\documentclass[submission, Phys,usenames,dvipsnames]{SciPost}
\usepackage[utf8]{inputenc}
\usepackage[T1]{fontenc}
\usepackage{lmodern}
\usepackage{amsmath,amssymb}
\usepackage{microtype}
\usepackage{array}
\usepackage{physics}
\usepackage{bbold}
\usepackage{xcolor}
\usepackage{listings}
\usepackage[mathletters]{ucs}
\usepackage{subfig}
\usepackage{caption}
\usepackage{comment}
\usepackage{soul}
\hypersetup{
    colorlinks,
    linkcolor={red!50!black},
    citecolor={blue!50!black},
    urlcolor={blue!80!black}
}

\usepackage[bitstream-charter]{mathdesign}
\DeclareSymbolFont{usualmathcal}{OMS}{cmsy}{m}{n}
\DeclareSymbolFontAlphabet{\mathcal}{usualmathcal}

\usepackage{inconsolata}
\lstdefinelanguage{Julia}%
  {morekeywords={abstract,break,case,catch,const,continue,do,else,elseif,%
      end,export,false,for,function,immutable,import,importall,if,in,%
      macro,module,otherwise,quote,return,switch,true,try,type,typealias,%
      using,while},%
   sensitive=true,%
   escapeinside={\%*}{*)},%
   morecomment=[l]\#,%
   morecomment=[n]{\#=}{=\#},%
   morestring=[s]{"}{"},%
   morestring=[m]{'}{'},%
}[keywords,comments,strings]%

\begin{document}
	\begin{center}{\Large \textbf{
	HarmonicBalance.jl: A Julia suite for nonlinear dynamics using harmonic balance
	}}\end{center}
	
	\begin{center}
        Jan Ko\v{s}ata\textsuperscript{1,*},
        Javier del Pino\textsuperscript{1,*},
        Toni L. Heugel\textsuperscript{1}, and
        Oded Zilberberg\textsuperscript{2}\\
    \end{center}
    
    \begin{center}
       {\bf 1} Institute for Theoretical Physics, ETH Z{\"u}rich, 8093 Z{\"u}rich, Switzerland\\
       {\bf 2} Department of Physics, University of Konstanz, 78464 Konstanz, Germany\\
       % TODO: provide email address of corresponding author
       ${}^\star$ {\small \sf  J. K. and J. d. P. contributed  equally to this work.}\\
       {\small \sf kosataj@phys.ethz.ch, jdelpino@phys.ethz.ch}
    \end{center}

    \begin{center}
    \today
    \end{center}
	
	%%%%%%%%%%%%%%%%%%%%%%%%%%%%%%%%%%%%%%%%%%%%%%%%%%%%%%%%%%%%%%%%%%%%%%%%%%%%%%%%%%%%%%%%%%%%%%%%%%%%%%%%%%%%%%%%%%%%%%%
	
	\begin{abstract}
     HarmonicBalance.jl is a publicly available Julia package designed to simplify and solve systems of periodic time-dependent nonlinear ordinary differential equations. Time dependence of the system parameters is treated with the harmonic balance method, which approximates the system's behaviour as a set of harmonic terms with slowly-varying amplitudes. Under this approximation, the set of all possible steady-state responses follows from the solution of a polynomial system. In HarmonicBalance.jl, we combine harmonic balance with contemporary implementations of symbolic algebra and the homotopy continuation method to numerically determine \textit{all} steady-state solutions and their associated fluctuation dynamics. For the exploration of involved steady-state topologies, we provide a simple graphical user interface, allowing for arbitrary solution observables and phase diagrams. HarmonicBalance.jl is a free software available at \url{https://github.com/NonlinearOscillations/HarmonicBalance.jl}.
	\end{abstract}
	
	\section{Introduction}
\begin{figure}
        \centering
        \includegraphics[width=0.7\textwidth]{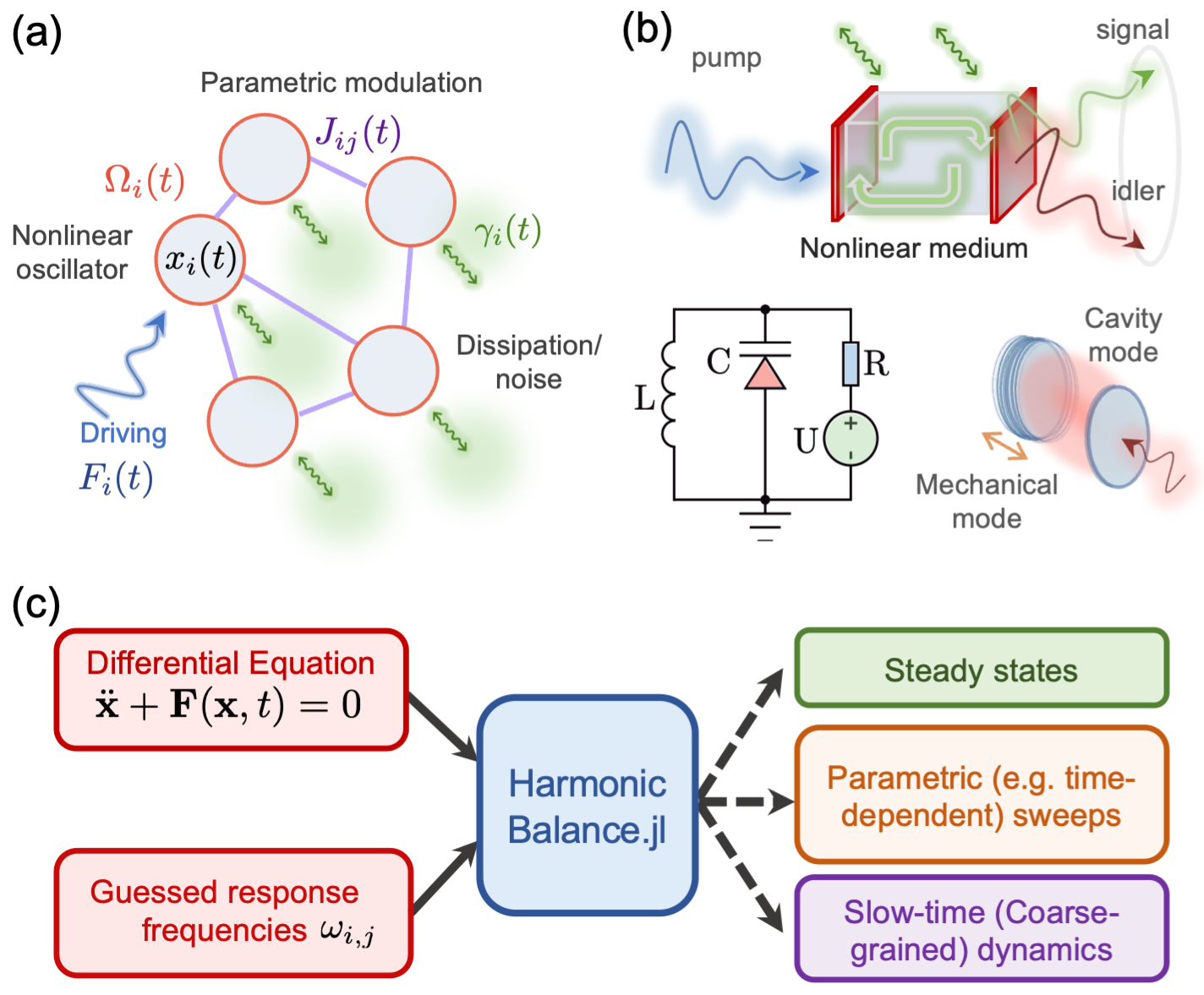}
        \caption{\textbf{Example systems treatable by HarmonicBalance.jl.} (a) A generic driven dissipative system: coupled nonlinear oscillators (circles) with amplitudes $x_i(t)$, time-varying natural frequencies $\Omega_i(t)$, and coupling amplitudes $J_{ij}(t)$ (here $i,j=1,2,\cdots,N$). System-environment interactions, parameterised by $\gamma_i(t)$, lead to fluctuations and dissipation. Harmonic drives with amplitudes $F_i(t)$ excite and modify the system's response. (b) Examples of nonlinear devices found in physics and engineering: (top) light fields interacting with nonlinear media, (bottom, left) driven nonlinear electric RLC circuits, and (bottom right) optomechanical resonators. (c) Basic functionality of HarmonicBalance.jl: A nonlinear (polynomial) ODE system with harmonic time-dependence is defined along with a set of expected response frequencies for each resonator, $\omega_{ij}$ (red input boxes). The package finds all dynamical steady states, identifies the most likely response given a parametric sweep, and supports numerical time-evolution within the harmonic ansatz (introduced in~\ref{sec:harm_exp}).}\label{fig:setup}
    \end{figure}

	Nonlinear ordinary differential equations (ODE) describe the time evolution of physical systems, in which distinct parts of the system interfere, cooperate or compete nonlinearly. Such ODEs host unique phenomena, including bifurcations, internal resonance, synchronisation, and chaos~\cite{vibrations1950jj,dykman1975theory,Strogatz1994,Rand_2005, nayfeh2008nonlinear,Schuster2009,Mangussi2016}. These manifest in a plethora of fields, such as fluid dynamics~\cite{Falkovich2011}, mechanics and robotics~\cite{niku2020introduction}, structural engineering~\cite{Ewins2000}, electronics~\cite{Rohde2005}, optics~\cite{Shen2002}, predator-prey dynamics~\cite{hofbauer1998evolutionary}, chemical oscillations~\cite{Epstein1999}, and biological processes~\cite{volkenstein2012molecular}. In particular, explicitly time-dependent or \textit{non-autonomous} nonlinear ODEs describe driven-dissipative systems, which commonly exhibit harmonic time dependence of system parameters and/or external drives~\cite{richards2012analysis}, see Fig.~\ref{fig:setup}a.
	
	Despite the ubiquitous presence of nonlinear dynamical systems in science and engineering [Fig.~\ref{fig:setup}b], their behaviour is not analytically tractable in most cases. Hence, one often resorts to using numerical ODE solvers e.g. as in molecular dynamics simulations~\cite{allen2017computer}.
	These usually focus on \textit{initial value problems}, where the system's state is advanced from a set of initial conditions. Linear systems, given sufficient time to freely evolve, usually relax to a unique \textit{stationary} or \textit{steady state}, i.e., a state where the system no longer evolves in time. %A harmonically driven/modulated system typically evolves towards a steady harmonic oscillating motion with constant amplitude and phase.
     In a harmonically driven system, the steady states will typically display time dynamics which are also harmonic, requiring a corresponding definition of a dynamical stationarity condition.
	
	Nonlinear systems challenge this approach since they can feature multiple steady-state solutions. The knowledge of \textit{all} such solutions is of key importance for analysing the system's behaviour, revealing experimentally observable phenomena such as hysteresis, spontaneous symmetry-breaking, or noise-induced switching dynamics~\cite{Leuch2016,2019PhRvL.123q3601H, Margiani2021}. However, when using a dynamical solver (or a time-resolved experiment), only a single steady state is found per run, depending on the initial conditions~\cite{Chitra2015,Roque2020}. Therefore, a complete exploration of the solution landscape would require infinite sampling of the continuous space of initial conditions. 
	
	An alternative approach to finding steady-state solutions in harmonic systems is transforming the system into a frame rotating with the applied drives~\cite{Arnold1978,cohen1986quantum}. In such a frame, the drives and corresponding steady states appear stationary, reducing the problem to finding the roots of a time-independent system of nonlinear equations~\cite{Bhaseen2012, Aldana2013}. While this approach is standard in few-variable problems, the proliferation of roots in multivariate nonlinear systems constitutes a challenge in itself~\cite{Bernshtein1975,sturmfels2002solving,cox2005solving}. Indeed, the most straightforward root-finding algorithms (e.g., the Newton-Raphson method) only find a single solution in the vicinity of an initial guess, making the exploration of all steady states infeasible. 
	
	Fortunately, many of the aforementioned nonlinear examples [cf. Fig.~\ref{fig:setup}b] feature ODE systems with polynomial dependence on the dynamical variables and their derivatives. For these, seeking all steady states in a given rotating frame generates coupled polynomial (algebraic) equations, which are numerically solvable using the \textit{homotopy continuation} method~\cite{Sommese2005,Breiding_2018,timme2021numerical}. This method utilises the continuous deformation of an exactly solvable problem into the problem of interest, thus finding \textit{all} the roots of coupled polynomials in one go. To our knowledge, however, no homotopy-continuation-based toolbox for the analysis of steady states of harmonically-driven systems has been developed so far.
	
    Here, we introduce HarmonicBalance.jl: an open-source suite that can find steady-state solutions to non-autonomous differential equations with harmonic time-dependencies. We unify a variety of existing methods for the analysis of time-dependent nonlinear ODEs into an integrated framework and take advantage of open-source Julia libraries to achieve ease of use, flexibility, and high performance [Fig.~\ref{fig:setup}c]. Our package is readily applicable to a range of active fields where nonlinear harmonically-driven systems appear, such as modal analysis in structural dynamics~\cite{Ewins2000,Kerschen2006}, electric circuits~\cite{Rohde2005,Rubiola2008,Fallis2003}, nonlinear optics~\cite{Haken1975,Shen2002,del2007optical,Rodriguez2016,DelPino2016,Sounas2018,Zambon_2020,peters2021limit}, optomechanics~\cite{Aspelmeyer2014,Pelka2020,Burgwal2020,Roque2020}, micro- and nanomechanics~\cite{Poot2012,Papariello2016,Guttinger2017,Chen2017,Houri2019,Houri2020,Yang_2021a, Yang_2021b, Mohammadi2020,Huber_2020,2018ApPhL.112w3105E,2020PhRvP..14a4042K,2021arXiv210911943H}, oscillator networks~\cite{2016PhRvA..93d3827O,2019PhRvL.123l4301H,Ozawa2019,2021Kosata,del2021non,2022arXiv220106315P},  Ising machines~\cite{Wang2013,Bello2019a,CalvaneseStrinati2019,2019PhRvL.123y4102F,CalvaneseStrinati2020,PhysRevResearch.4.013149}, and many-body light-matter systems~\cite{griffin1996bose,Carusotto2013,Soriente2018,Kirton2019,Soriente2020,Soriente2021,2021PhRvX..11d1046F}. 
     
	\section{Harmonically-driven nonlinear systems: basic principles}\label{sec:model}
	
	HarmonicBalance.jl focuses on harmonically-driven nonlinear systems, i.e., dynamical systems governed by equations of motion where all explicitly time-dependent terms are harmonic. In the exposition here, we will assume a general nonlinear system of $N$ second-order ODEs\footnote{Second-order ODEs based on the harmonic oscillator represent the vast majority of expected use cases. However, the methods described in this work are applicable to arbitrary ODE orders~\cite{richards2012analysis}.} with real variables $x_i(t)$, with index $i = 1,2,\cdots,N$ and time $t$ as the independent variable,
	\begin{equation} \label{eq:ode}
	  \ddot{\vb{x}}(t)+ \vb{F}(\vb{x}(t), t)=0\:.
	 \end{equation}
	 The vector $\vb{x}(t) = (x_1(t), ..., x_N(t))^{\text T}$ fully describes the state of the system.  Physically, $\vb{x}(t)$ encompasses the amplitudes of either point-like or collective oscillators (e.g., mechanical resonators, voltage oscillations in $RLC$ circuits, an oscillating electrical dipole moment, or standing modes of an optical cavity). We assume $\textbf{F}(\textbf{x}(t),t)$ can be decomposed into a sum of $L$ periodic terms\footnote{We omit the time-dependence of $x(t)$ where this is clear from the context.}:,
	 \begin{equation}
	  \vb{F}(\vb{x}, t) = \vb{f}_0(\vb{x}) + \sum_{l=1}^L \vb{f}_{l}(\vb{x}) \cos(\omega_l t + \phi_l)\,,\label{eq:F_xt}
	\end{equation}
	with vector fields $\vb{f}_l(\vb{x})$.  The field $\vb{f}_0$ describes static properties of the system while $\vb{f}_{l\neq0}$ account for explicit time-dependence, typically induced by one or more sources of periodic driving and/or parameter modulation with frequencies $\{\omega_l\}$ and phases $\{\phi_l\}$. In Table~\ref{table:terms}, we list several examples of terms commonly constituting Eq.~\eqref{eq:F_xt}. 
	
	\begin{table}[t] 
	    \centering
        \begin{tabular}{ |c|c|c| } 
         \hline
         term in $F_i(\vb{x}, t)$ & physical mechanism & frequency conversion \\ \hline
         $x_i$ & natural response (spring constant) & - \\
         $x_j$ & mode coupling & - \\
         $\dot{x}_i$ & damping/gain & -\\
         $\dot{x}_j$ & dissipative coupling & -\\
         $\cos(\omega_d t)$ & external drive (frequency $\omega_d$) & -\\
         $x_i^2$ & Pockels coefficient & $\omega_i \rightarrow 2\omega_i$ \\ 
         $x_i^3$ & Kerr (Duffing) coefficient & $\omega_i \rightarrow 3 \omega_i$ \\ 
         $x_i^2 \dot{x}_i$ & nonlinear damping & $\omega_i \rightarrow 3 \omega_i$\\ 
         $x_i \cos(\omega_{\rm p} t)$ & parametric drive (frequency $\omega_{\rm p}$) & $\omega_i \rightarrow \omega_i \pm \omega_{\rm p}$\\
         $x_j x_i $ & nonlinear mode interactions & $\omega_i \rightarrow \omega_i \pm \omega_j$\\ 
         \hline
        \end{tabular}
	     \caption{\textbf{Examples of terms occurring in Eq.~\eqref{eq:F_xt} and their origins.} The rightmost column shows the frequency conversion taking place, assuming two variables $x_i$, $x_j$, oscillating at frequencies $\omega_i$, $\omega_j$ (for example, $\omega_i \rightarrow 2\omega_i$ means, that oscillating at frequency $\omega_i$ leads to additional oscillations at frequency $2\omega_i$).}
	     \label{table:terms}
	\end{table}
	
	Note that in a linear system, $\vb{F}(\vb{x}, t)$ can be written as $\vb{F}(\vb{x}, t)=\vb{M}(t) \vb{x}+ \textbf{b}(t)$, where the matrix $\vb{M}(t)$ contains spring constants and linear couplings, while $\vb{b}(t)$ is a vector of external forces. Diagonalising $\vb{M}(t)$ then yields the so-called \textit{normal modes} of the system. While the notion of normal modes does not directly apply in a nonlinear system, they usually constitute a convenient basis choice for a perturbative treatment.

	% NOTE: our system is not actually periodic! (drives can be incommensurate)
	\subsection{The harmonic expansion}\label{sec:harm_exp}
	
	For sufficiently long times (i.e.,~after any transient responses have disappeared), the solutions 
	of Eq.~\eqref{eq:ode} are expected to appear as a sum over harmonics. Let us illustrate this point using the example of driven harmonic oscillators. In this simple case, Eqs.~\eqref{eq:ode} and \eqref{eq:F_xt} take the form
	\begin{equation} \label{eq:shos}
	\ddot{\vb{x}}(t) + \vb{M} \vb{x}(t) = \vb{g} \cos(\omega_d t)\,,
	\end{equation}
	where we assumed a constant vector $\vb{g}$ and drive frequency $\omega_d$.	The standard method to solve for the steady states is to Fourier-transform both sides of Eq.~\eqref{eq:shos}; the resulting equations give an exact solution for $\vb{x}(t)$ in terms of its Fourier coefficients,
	\begin{equation} \label{eq:ft_x}
	    \tilde{\vb{x}}(\omega) = (\vb{M} - \omega^2 \mathbb{1})^{-1} \vb{g} \left[\delta(\omega + \omega_d) + \delta(\omega - \omega_d)\right] / 2\;,
	\end{equation}
	where $\tilde{\vb{x}}(\omega) = \int_{-\infty}^{+\infty} \vb{x}(t) e^{i\omega t}\: dt$ and $\delta(z-z_0)$ is the Dirac delta function of $z$ centred at $z_0$. This procedure is effective because in the Fourier domain, the l.h.s.~of Eq.~\eqref{eq:shos} becomes diagonal (i.e., it involves a single frequency) while the applied drive reduces to a Dirac delta function centred at $\omega=\omega_d$. We can thus select a single frequency out of the continuous space of all frequencies for which the solution is nonvanishing. Furthermore, due to the linear superposition principle, responses to arbitrary driving terms can be constructed out of the solution in Eq.~\eqref{eq:ft_x}.

	The same task becomes intractable if nonlinear terms are introduced. Nonlinearities facilitate \textit{frequency conversion} by coupling different harmonics of the system, rendering $\vb{F}(\vb{x},t)$ non-diagonal in Fourier space. As an example, let us consider a single driven Duffing oscillator~\cite{Lifshitz_2008}, whose equation of motion reads
	\begin{equation} \label{eq:duff_basic}
	    \ddot{x}(t) + \omega_0^2 x(t) + \alpha x^3(t) = F \cos(\omega_d t) \:.
	\end{equation}
	where $\omega_0$ is the natural frequency, $\alpha$ is the nonlinear coefficient, $F$ is the drive amplitude, and $x$ is now a scalar.
	The nonlinear (Duffing) term in Fourier space reads
	\begin{equation} \label{eq:duffingFT}
	    \alpha \int x^3(t) e^{-i\omega t} \: dt = \alpha \int_{-\infty}^{+\infty} \tilde{x}(\omega')\tilde{x}(\omega'')\tilde{x}(\omega''') \delta(\omega'''+\omega''+\omega'-\omega) \: d\omega' \: d\omega'' \: d\omega''' \,,
	\end{equation}
	coupling all combinations of four harmonics that sum to zero. This results in frequency-conversion processes known as four-wave mixing, which here convert, to lowest order, the driven oscillation at frequency $\omega_d$ to frequency $3\omega_d$; four-wave mixing appears as off-diagonal terms in Fourier space. The frequency conversion subsequently propagates through the entire spectrum, generating an infinite number of Fourier components. A nonlinearity thus precludes a closed-form solution of the problem in Fourier space; this is a common trait of nonlinear physical systems; see Table~\ref{table:terms} for examples of frequency-converting effects.
	
	A serviceable approach to numerically approximate the harmonics of a driven nonlinear system involves truncating the spectrum  $\tilde{\vb{x}}(\omega)$ to a finite set of frequencies~\cite{Guckenheimer_2013}. The idea of truncation in Fourier space is at the core of several widely-used methods, such as harmonic balance~\cite{Krack_2019, Luo_2012}, the rotating-wave approximation~\cite{Gardiner2004}, the van der Pol transformation~\cite{Rand_2005} in combination with Krylov-Bogoliubov averaging~\cite{Nayfeh_2008}, Magnus expansion~\cite{Mikami_2016, Eckardt_2015}, secular perturbation theory~\cite{Lifshitz_2008}, and also appears in the contemporary concept of Floquet engineering~\cite{Rudner_2020, Goldman_2014}. We implement the approach of harmonic balance using a generalised ansatz~\cite{Sarrouy_2011, Guskov_2007}
	\begin{equation} \label{eq:ansatz}
	     x_i(t) = \sum_{j=1}^M u_{i,j}  (T)  \cos(\omega_{i,j} t)+ v_{i,j} (T) \sin(\omega_{i,j} t) \:,
	\end{equation}
	where the sum runs over the finite set of desired frequencies $\{\omega_{i,j}\}$ describing the coordinate $x_i$. Here, $T$ represents a "coarse-grained" timescale that is much slower than the oscillations in the system ($T\gg 2\pi/\min \{\omega_j\}$).
	Equation~\eqref{eq:ansatz} represents an attempt to capture the dynamics of the system using a discrete set of real functions $\{u_{i,j}(T),v_{i,j}(T)\}$, which in our package we dub the \textit{harmonic variables}, while the fast oscillations are accounted for by the sine/cosine terms. Note that this ansatz would be exact if the set of frequencies $\{\omega_{i,j}\}$ covered all real numbers. Generally, it is not straightforward to identify the relevant frequencies $\{\omega_{i,j}\}$. A good starting point is the frequencies of any external drives, combined with the frequency conversions presented in Table~\ref{table:terms}. Alternatively, one can pick the highest-weight discrete Fourier components in a time trace obtained numerically or from experimental data.
	
	Plugging the ansatz Eq.~\eqref{eq:ansatz} into the ODE~\eqref{eq:ode}, we obtain equations governing pairs of harmonic variables $(u_{i,j}, v_{i,j})$ by convolving both sides with $\cos(\omega_{i,j} t)$ and $\sin(\omega_{i,j} t)$, respectively (see Appendix~\ref{sec:slow_dif}). The harmonic variables themselves are treated as constants during this step.
	We thus obtain two equations for each $\omega_{i,j}$. To reflect the slowly-changing nature of the harmonic variables, we drop all of their time derivatives of order two or higher, e.g., $\ddot{u}_{i,j}, \ddot{v}_{i,j}\rightarrow 0$. The resulting set of equations has no explicit time dependence; we call these the \textit{harmonic equations},
	\begin{equation} \label{eq:harmoniceq}
	    \frac{d\vb{u}(T)}{dT}  = \bar{\vb{F}} (\vb{u})\,,
	\end{equation}
	where $\vb{u}(T) = (u_{1,1}(T), v_{1,1}(T), ...,u_{N,M}(T), v_{N,M}(T))^{\text T}$ contains the harmonic variables and $\bar{\vb{F}} (\vb{u})$ the corresponding Fourier components resulting from inserting the truncated ansatz Eq.~\eqref{eq:ansatz} into Eq.~\eqref{eq:F_xt}. Note that Eq.~\eqref{eq:harmoniceq} still describes a slow time dependence of $\{u_{i,j}(T), v_{i,j}(T)\}$, or in other words, Eq.~\eqref{eq:harmoniceq} captures an  intrinsic frequency bandwidth $\Delta\omega \ll \omega_{i,j}$ of the response around the expansion frequencies $\{\omega_{i,j}\}$  in a similar spirit as the Krylov-Bogoliubov averaging~\cite{Nayfeh_2008,Leuch2016} and the rotating-wave-approximation~\cite{Gardiner2004}. Note that our scheme scales up very fast - each harmonic $\omega_{i,j}$ of a variable $x_i$ is converted into two harmonic variables. Hence, for a system of $N$ interacting components, each expanded in $M$ harmonics, Eq.~\eqref{eq:harmoniceq} consists of $2 N M$ harmonic equations. Note that the discussed approach is not exclusive to fixed points, and the ansatz can account for periodic asymptotic orbits, i.e., to limit cycles~\cite{vibrations1950jj,dykman1975theory,Strogatz1994,Rand_2005, nayfeh2008nonlinear,CarlonZambon2020,Ganesan2017,Czaplewski2018,Dykman2019}. Limit cycles can emerge when the system loses stability around Hopf bifurcations and a self-sustained oscillation emerges. We plan to include limit cycle functionality in future releases of the package.

	\subsection{Solving the harmonic equations for steady-state solutions}\label{sec:harmonic_eq_solving} 
	There are, in principle, two ways to extract the long-time behaviour of the system from Eq.~\eqref{eq:harmoniceq}. First, one can make use of an initial-value ODE solver to numerically propagate a state in $T$, starting at $\vb{u}(T=T_0)$. Given sufficient time, the system will typically converge towards a steady state $\vb{u}_0$. Secondly, we can look for steady states by explicitly requiring the l.h.s. of Eq.~\eqref{eq:harmoniceq} to vanish, leaving us with the task of solving the nonlinear \textit{algebraic} equations system
	\begin{equation}\label{eq:harmonic_balance}
	    \bar{\vb{F}} (\vb{u}_0) = 0\:.
	\end{equation} 
	The second approach is advantageous since, in principle, all steady states of the system can be found, irrespective of whether or how they can be reached by time evolution. In general, however, obtaining the roots of coupled nonlinear algebraic equations is highly non-trivial: for $n$ coupled polynomial equations of order $p$, B\'{e}zout's theorem~\cite{Cox_2013} provides an upper bound $p^n$ for the number of distinct complex solutions. For example, in the Duffing oscillator shown in Eq.~\eqref{eq:duff_basic}, we have one variable $x$ - given an ansatz Eq.~\eqref{eq:ansatz} with a single harmonic. This generates 2 polynomial harmonic equations of order 3, leading to a maximum of $3^2=9$ solutions. Generally, for $N$ coupled Duffing oscillators each expanded in $M$ harmonics, we have $2NM$ harmonic equations of order 3, leading to an upper bound of $3^{2NM}$ solutions. The exponential scaling rapidly makes charting the complete steady-state solution landscape
	a formidable challenge. Accordingly, many applications of harmonic balance deal with few-variable systems and rely on the perturbative treatment of nonlinearities.
	
	\begin{figure}[t]
        \centering
        \includegraphics[width=\textwidth]{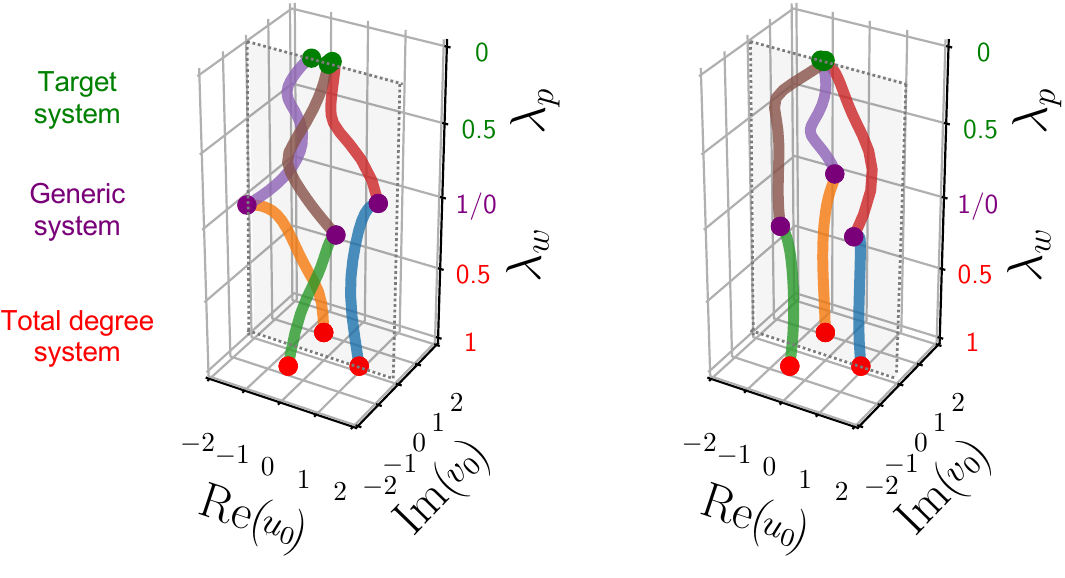}
        \caption{\textbf{Homotopy continuation in a Duffing resonator [cf.~Eq.~\eqref{eq:duff_basic}]} Path tracking from the complex roots of the system $\{u_0^3-1=0,v_0^3-1=0\}$ (red dots) to the roots of the harmonic balance equations [cf.~Eq.~\eqref{eq:harmonic_balance}, Appendix~\ref{sec:slow_dif}] (green dots), denoted $\vb{u}_0=(u_0,v_0)$, using homotopy continuation.  In the algorithm to track multiple parameter steady states, we carry out a two-step process encompassing a warm-up tracking (deformation parameter $\lambda_w\in(0,1)$), where singular paths (not-shown) are filtered-out, followed by a subsequent tracking to the harmonic balance equations, (deformation parameter $\lambda_p\in(0,1)$) [see Appendix~\ref{sec:solving_multiple} for further details].  The roots $\vb{u}_0$ traverse the complex plane until paths intersect with the planes $\mathrm{Im}(u_0)=0$ and $\mathrm{Im}(v_0)=0$ (light gray), at the 3 real  roots of $\vb{F}(\vb{u}_0)$. For this figure $\alpha=1$, $\omega_d=1.03$, $\omega_0=1$, $F=0.01$, $\theta=0$, and $\gamma=0.01$.}
        \label{fig:path_track}
    \end{figure} 
	
	In our package, we solve the algebraic Eq.~\eqref{eq:harmonic_balance} using the homotopy continuation method~\cite{Bates_2013, Verschelde_1999} as implemented by the open-source package HomotopyContinuation.jl~\cite{Breiding_2018}. To find the roots of a polynomial, this method, in its simplest form, starts from another analytically-solvable polynomial of the same order, which saturates the Bézout bound. For instance, to find the roots of a single, $p$-th order polynomial $P(z)$ of the variable $z$, one can start from the polynomial $U(z)=z^p - 1$ with roots $e^{2\pi i k /p}$ ($k = 1,..., p$). Employing this so-called \textit{total-degree homotopy}, the polynomial $U(z)$ is "slowly" deformed~\footnote{Namely, via the homotopy $H(U(z),P(z),\lambda)=\lambda U(z)+(1-\lambda)P(z)$, with deformation parameter $\lambda$. In the current context, the homotopy is a mapping such that $H(U(z),P(z),\lambda=1)=U(z)$ and $H(U(z),P(z),\lambda=0)=P(z)$.} into the polynomial $P(z)$, correcting the known roots (i.e., tracking the roots) after each step. At the end of the homotopy continuation, the obtained set of roots is guaranteed to be complete, as it has been tracked from the full set of $p$ complex roots of $U(z)$.
	
	A generalisation of the total degree homotopy approach enables the solution of ~Eq.~\eqref{eq:harmonic_balance}, by instead tracking the roots of the uncoupled system  $\vb{U}(\vb{u}_0)=\{u_1^{d_1} - 1,v_1^{d_2} - 1,u_2^{d_3} - 1,v_2^{d_4} - 1,\cdots\}$, with $d_r$ equal to the degree of the polynomial $\bar{F}_r(\vb{u}_0)$, to those of $\bar{\vb{F}}(\vb{u}_0)$ (here $r=1,2,\cdots,2NM$ is an index labelling all harmonic balance equations). Such an approach leads to a number of solution paths to track that scales exponentially with $NM$. This constitutes a challenge in usual exploratory research scenarios that seek to find the behaviour of the solutions as parameters vary. Crucially, in physical problems (and in contrast with single polynomials), $\bar{\vb{F}}(\vb{u}_0)$ often displays a vastly smaller number of roots than the B\'{e}zout bound (see Examples). This implies that many solution paths from $\vb{U}(\vb{u}_0)=0$ are usually irrelevant (singular\footnote{A solution $\vb{u}_0$ is denoted singular if the Jacobian matrix for $\bar{\vb{F}}(\vb{u})$ at $\vb{u}=\vb{u}_0$ has full rank. These solutions are associated with the intersection, reversal or divergences of solution paths, and their early detection is crucial in efficient path tracking.}) since they do not lead to roots of $\bar{\vb{F}}(\vb{u}_0)$. In HarmonicBalance.jl, we harness a common strategy to reduce the computational overhead in multi-parameter root finding~\cite{Sommese2005}. This strategy consists of the two-step algorithm described in Appendix~\ref{sec:solving_multiple} and exemplified in~Fig.~\ref{fig:path_track}.
	
	Finally, note that although the obtained solutions are complex, only real roots of~Eq.\eqref{eq:harmonic_balance} are physically meaningful\footnote{Complex roots with small imaginary parts suggest that small perturbations in the problem might introduce a new real root, see Ref.~\cite{Sommese2005}.}. For a more detailed description of the method and its implementation, see the documentation of HomotopyContinuation.jl and references therein~\cite{Breiding_2018, Sommese2005}. 
	
	\subsection{Stability analysis and linear response} \label{sec:stability_analysis} 
	Let us assume that we found a real solution $\vb{u}_0$ of Eq.~\eqref{eq:harmoniceq}. When the system is in this state, it responds to small perturbations either by returning to $\vb{u}_0$ over some characteristic timescale (\textit{stable state}) or by evolving away from $\vb{u}_0$ (\textit{unstable state}). To analyze the stability of $\vb{u}_0$, we linearize Eq.~\eqref{eq:harmoniceq} around $\vb{u}_0$ for a small perturbation $\delta \vb{u} = \vb{u} - \vb{u}_0$ to obtain
	\begin{equation} \label{eq:Jacobeq} 
	    \frac{d}{dT} \left[\delta \vb{u}(T)\right] =  J(\vb{u}=\vb{u}_0) \delta \vb{u}(T) \:,
	\end{equation}
	where $J(\vb{u}_0)=\grad_{\vb{u}}  \bar{\vb{F}}|_{\vb{u}=\vb{u}_0}$ is the \textit{Jacobian matrix} of the system evaluated at $\vb{u}=\vb{u}_0$\footnote{In this notation, $\grad_{\vb{u}}  \bar{\vb{F}}=(\grad^{\text T}_{\vb{u}}\bar{F}_1,\grad^{\text T}_{\vb{u}}\bar{F}_2,\cdots, \grad^{\text T}_{\vb{u}}\bar{F}_{2NM})^{\text T}$, where $\grad^{\text T}_{\vb{u}}\bar{F}_r$ is the transpose (row vector) of the gradient of the $r$-th component over the coordinates $\vb{u}$.}.
	
	The solution to Eq.~\eqref{eq:Jacobeq} can be expanded in terms of the complex eigenvalues $\lambda_r$ and eigenvectors $\vb{v}_r$ of $J(\vb{u}_0)$, namely
	\begin{equation}\label{eq:fluct_evo}
	    \delta \vb{u}(T) = \sum_{r=1}^{2NM}(\vb{v}_r\cdot \delta\vb{u}(T=T_0))\hspace{1mm}\vb{v}_r e^{\lambda_r T}.
	\end{equation}
	The dynamical behaviour near the steady states is thus governed by $e^{ \lambda_r T}$: if $\mathrm{Re}(\lambda_r)<0$ for all $r$, the system returns to $\vb{u}_0$ under a small arbitrary perturbation and the state is classified as stable. Conversely, if $\mathrm{Re}(\lambda_r)>0$ for any $r$, the system is unstable - perturbations such as noise or a small applied drive will force it to evolve away from $\vb{u}_0$. 

	 As the system is coupled to dissipative baths, it will be also subject to fluctuations. While the current focus of our package is deterministic dynamics, analysis of weak fluctuations around a stable fixed point is included. The linear response of a stable, steady state to an additional oscillatory force, caused by weak probes or noise, is often observed in experiments~\cite{Dykman2012}. It can be calculated by adding a small driving term $\vb{\delta f} \cos(\Omega_d T)$ to the harmonic Eq.~\eqref{eq:harmoniceq}. To account for the time dependence of the perturbation, linearisation around $\vb{u}_0$ should retain the previously-dropped higher-order time derivatives. While implemented in the package, we leave a thorough discussion of this topic, as well as noise-activated dynamics, to future work. 
	 
	 %For details on the calculation of the linear response, see Section~\ref{sec:Jacobians}.
	 %\begin{equation} \label{eq:linresponse}
	 %    (\grad_{\vb{u}}  \vb{F})|_{\vb{u}=\vb{u}_0} \vb{\delta u}(T) +  (\grad_{\dot{\vb{u}}} \vb{F})|_{\vb{u}=\vb{u}_0}  \delta %\dot{\vb{u}}(T) +  (\grad_{\ddot{\vb{u}}} \vb{F} )|_{\vb{u}=\vb{u}_0}\delta \ddot{\vb{u}}(T) = \vb{\delta f} \cos(\Omega T)\:.
	 %\end{equation}
	 
	 %Assuming harmonic behaviour of $\vb{\delta u}(T)$, Eq.~\eqref{eq:linresponse} reduces to a linear system of equations. Note that $\vb{\delta f}$ describes the force acting on the harmonic variables $\vb{ \vb{u}}$; in Appendix~\ref{sec:Jacobians}, we describe the relation between $\vb{\delta f}$ and forces in the untransformed equations of motion.

	\begin{figure}
        \centering
        \includegraphics[width=150mm]{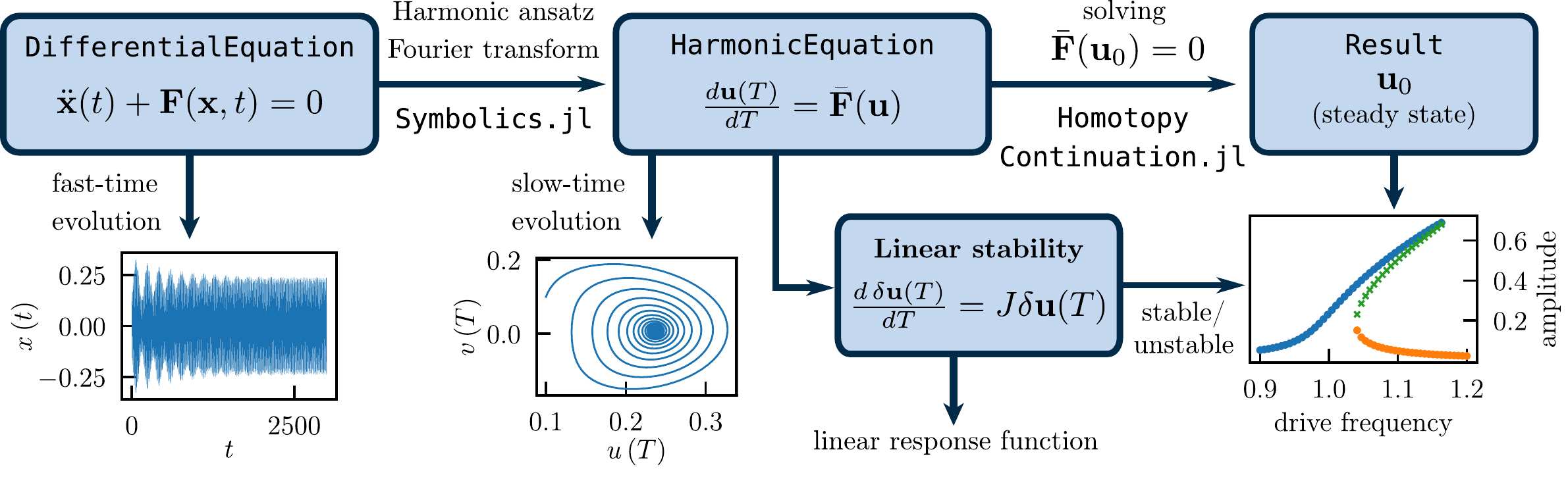}
        \caption{The basic workflow of HarmonicBalance.jl.}
        \label{fig:chart}
    \end{figure} 
    
\section{Structure of HarmonicBalance.jl}\label{eq:structure}

The bulk of HarmonicBalance.jl is written in Julia, a language combining the accessibility of interpreted languages such as Python with the performance of compiled languages such as C and Fortran. Our package is designed with simplicity of use and scalability in mind. We give a short overview of the package structure and basic workflow below; it is also displayed in Fig.~\ref{fig:chart}.
    \begin{enumerate}
        \item ODE systems serve as primary input and are inserted in symbolic form via Symbolics.jl. Non-autonomous harmonic systems are accompanied by a user-defined set of frequencies to build the ansatz [cf.,~Eq.\eqref{eq:ansatz} and Fig.~\ref{fig:setup}c]. From this, a set of harmonic (autonomous) equations is obtained symbolically.
        \item Harmonic equations are interfaced with a root-finding algorithm (HomotopyContinuation.jl) and numerical ODE solvers (DifferentialEquations.jl).
        \item  Steady-state solutions are further analysed, e.g., for stability and other criteria. A flexible toolbox for the calculation of observable quantities is included, with a user-defined mapping of steady-state solutions and the calculation of linear response spectra. 
        \item Steady-state solutions for one- and two-dimensional parameter sets are readily visualised. Interactive plotting routines help exploratory analysis of the rich topology of the solutions (e.g., bifurcations) for multi-dimensional parameter sets. An ODE solver may be used to obtain the slow-time ($T$) dynamics and verify the steady-state results.
    \end{enumerate}

    In the following, we detail the working principles of HarmonicBalance.jl. For detailed online documentation on the latest release, we refer the reader to~\cite{harmonic_balance_docs}.

    \subsection{Defining a system, extracting harmonic equations.}
    
    Conceptually, specifying a system requires two ingredients: its ODE of motion, such as Eq.~\eqref{eq:ode}, and the set of oscillating frequencies forming the harmonic ansatz in Eq.~\eqref{eq:ansatz}. Once defined, the equation of motion is stored in the dedicated object \texttt{DifferentialEquation}. For this primary input and subsequent symbolic manipulations, we employ the Julia package Symbolics.jl~\cite{10.1145/3511528.3511535}, whose emphasis on high performance is essential in dealing with complex problems, such as that shown in Section~\ref{sec:hardexample}. 
    
     After constructing the harmonic ansatz Eq.~\eqref{eq:ansatz} and introducing the slow time $T$, the harmonic equations governing the harmonic variables $\vb{u}$ are found. We assume slow or no evolution in $T$ and therefore drop any second and higher-order derivatives with respect to $T$ as well as products of $\frac{d \vb{u}}{dT}$. The equations are then Fourier-transformed, returning the harmonic Eq.~\eqref{eq:harmoniceq} (for a concrete example, see Appendix~\ref{sec:slow_dif} and Section~\ref{sec:duffing_example}). These equations are stored in the object \texttt{HarmonicEquation}, which in itself stores instances of \texttt{HarmonicVariable}, each specifying one pair $\{u_{i,j}, v_{i,j}\}$.

    \subsection{Obtaining and characterising steady states}
    
    To find and analyse the steady states of a \texttt{HarmonicEquation}, we need the corresponding (algebraic) steady-state equations and the Jacobian matrix; these are symbolically stored in the object \texttt{Problem}.
    
    As the next step, numerical parameter values are specified. This is the required input for our primary solving method, \texttt{get\_steady\_states}, which allows the user to specify which parameters are constant and which are varied. One may vary any number of parameters, i.e., solution sets of any dimension are supported. We then employ a homotopy to retrieve steady states for each parameter set value. The native multi-threading support of HomotopyContinuation.jl can dispatch path tracking over multiple cores. Note that complex roots are being followed throughout the procedure - real solutions are only filtered out a posteriori.
    
    Once steady states are found, HarmonicBalance.jl automatically classifies each state by whether it is real (complex roots bear no physical meaning) and by its stability. The final output is a \texttt{Result} object.
    
    \subsection{Visualisation}\label{subsec:vis}
    
    Functionality for static plotting of steady-state and time-dependent solutions is provided using the \texttt{Matplotlib} library~\cite{Hunter2007}. At the core of our plotting routines is the function \texttt{transform\_solutions}, which evaluates a symbolic expression by substituting each of the solutions from a \texttt{Result} object. This enables 1D and 2D plots of each solution and functions derived from them. Functionality for selection of transformed solutions (e.g. those with the maximal amplitudes) is available through the method \texttt{map\_multi\_solutions}\footnote{In general, this function provides support for many-to-one functions of multiple transformed solutions.}. We implemented classification functionality to help navigate potentially involved solution landscapes to show/hide/label solutions satisfying a given condition.
    
    Rather than the solutions themselves, a \textit{phase diagram} is often desired, which shows how the qualitative behaviour of the system changes in parameter space. We provide the functionality to distinguish parameter space regions by the number of (stable/all) solutions. We include an additional test that is able to detect qualitative differences between points in parameter space where the number of stable/unstable solutions is invariant (e.g. transcritical bifurcations)\footnote{In short, this function first labels each point in parameter space with a bit string indicating the stability of each solution there (e.g. \texttt{[is\_stable(sol\_1),is\_stable(sol\_2),$\cdots$ ]=[011$\cdots$])}. Next, each bit string is assigned to a unique integer identifier.}.
    
    Finally, we highlight the function \texttt{interactive}$\_$\texttt{phase}$\_$\texttt{diagram}$\_$\texttt{2D}; this produces an interactive window where the solution landscape can be explored along a given dimension by simply clicking on different parts of a two-dimensional diagram. 
    
    \subsection{Time-dependent simulations}\label{sec:time_dep}
    
    The behaviour of a system can be found using a numerical ODE solver and an initial condition $\vb{u}(T=T_0)$. This approach is beneficial for analysing systems whose parameters are being adiabatically varied in time, i.e. slower than characteristic system reaction times - a common experimental approach to explore the solution landscape, detecting bifurcations and hysteresis. Other potential uses of time-dependent simulations are verifying steady-state results, their stability and fluctuation spectra, and identifying their basins of attraction.
    
    A time-dependent simulation may use either a \texttt{DifferentialEquation} or a \texttt{HarmonicEquation}. The former represents the system in the time domain and uses no approximations. However, the numerical propagation of oscillatory dynamics can be extremely inefficient. Specifically, time grids need to resolve fractions of the period of the fastest oscillation to keep numerical precision. A \texttt{HarmonicEquation} is significantly faster to solve since it describes the system using the slow time $T$, where the oscillatory steady states appear time-independent. This approach, however, can only capture the chosen set of harmonics and nearby frequencies through the slowly varying amplitudes $\vb{u}$. To include additional frequencies, the harmonic ansatz must be expanded. To illustrate the interface to time-dependent solvers, we provide a simple example in the online documentation \url{https://nonlinearoscillations.github.io/HarmonicBalance.jl/stable/examples/time_dependent/}.

    \section{Comparison with other harmonic balance implementations}
    
    Steady-state problems in nonlinear periodic ODEs appear in diverse areas of science and technology. Several free and commercial packages exist to solve them. Examples of open-source software include Xyce - a high-performance parallel electronic simulator that can perform harmonic balance analysis~\cite{verley2018xyce}. The harmonic balance method is also natively supported for general nonlinear multiphysics finite element simulations in the open-source C++ FEM library Sparselizard~\cite{halbach2017sparselizard}. An example of a commercial software package is Cadence AWR Microwave Office, which focuses on the analysis of electrical circuits in the frequency domain (i.e. calculation of voltage and current for a given set of elements), and includes features that are not yet implemented in HarmonicBalance.jl (e.g., noise analysis). Generally, these packages are use-case specialised, making their application in other settings challenging.
    
    Likely the closest existing tool to our Julia package is NLvib~\cite{Krack_2019}. However, it focuses
    on predefined problems primarily relevant to mechanical engineering and requires a MATLAB license to use, which is unaffordable for many research groups and educational institutes. Crucially, in all of the above packages, the harmonic balance equations are solved by either \textit{i)} time evolution from a given set of initial conditions or \textit{ii)} single-root finding methods such as Newton's. These approaches only return one steady state at a time, and even when combined with the continuation of a known solution, disconnected solution branches remain hidden. Our distinction from existing work lies in three main points:
    \begin{itemize}
        \item The use of homotopy continuation allows us to find all possible solutions of the harmonic balance equations, where solutions for multiple parameters can be obtained reliably and efficiently. In this critical step, usually the most computationally intensive, we rely on HomotopyContinuation.jl, a package which outperforms other established homotopy continuation libraries~\cite{Breiding_2018}.
        \item Arbitrary equations of motion can be entered and processed. This is crucial for use by the academic community, where an ab-initio approach to physical problems is often preferred over specialised GUI-based tools.
        \item The code and its dependencies are entirely open-source. This enables a natural synergy with Julia's existing rich ecosystem for scientific computing.
    \end{itemize}
	
	\section{Examples of HarmonicBalance.jl usage} \label{sec:examples}
	In this Section, we present several example problems. Instructions to install HarmonicBalance.jl and further detailed examples can be found at \url{https://github.com/NonlinearOscillations/HarmonicBalance.jl} and in the links therein.
	\subsection{Duffing oscillator} \label{sec:duffing_example}
	
	\begin{figure}[h]
        \centering
        
        \subfloat[I. Characteristics of the dominant harmonic ($\omega_d\rightarrow \omega_d\approx\omega_0$) in the steady state.]{\includegraphics[width=\textwidth]{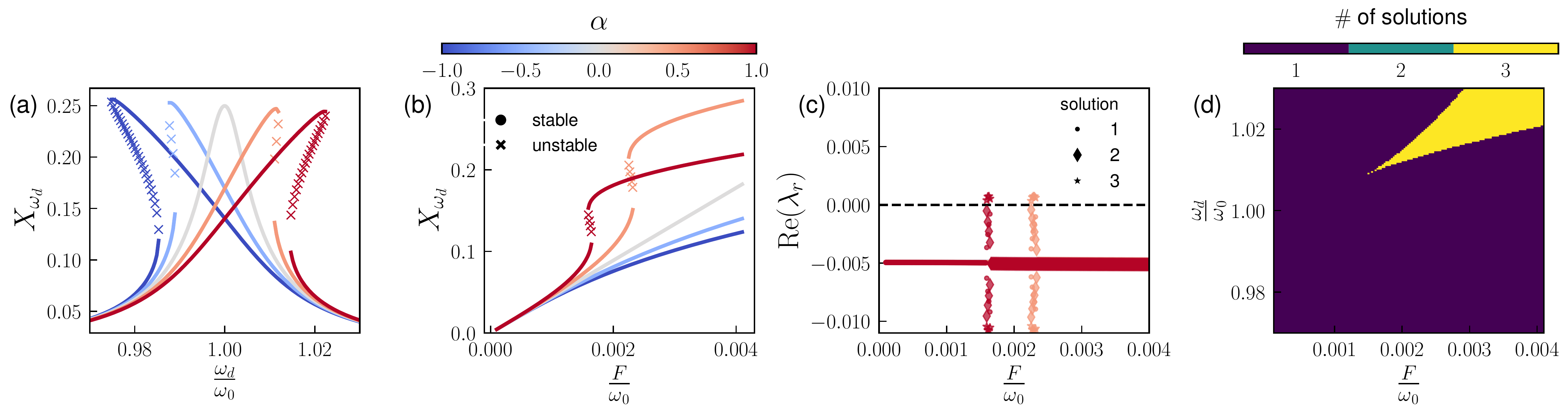}}
        
        \subfloat[II. Characteristics of the superharmonic response ($\omega_d\rightarrow 3\omega_d\approx\omega_0$) in the steady state.]{\includegraphics[width=\textwidth]{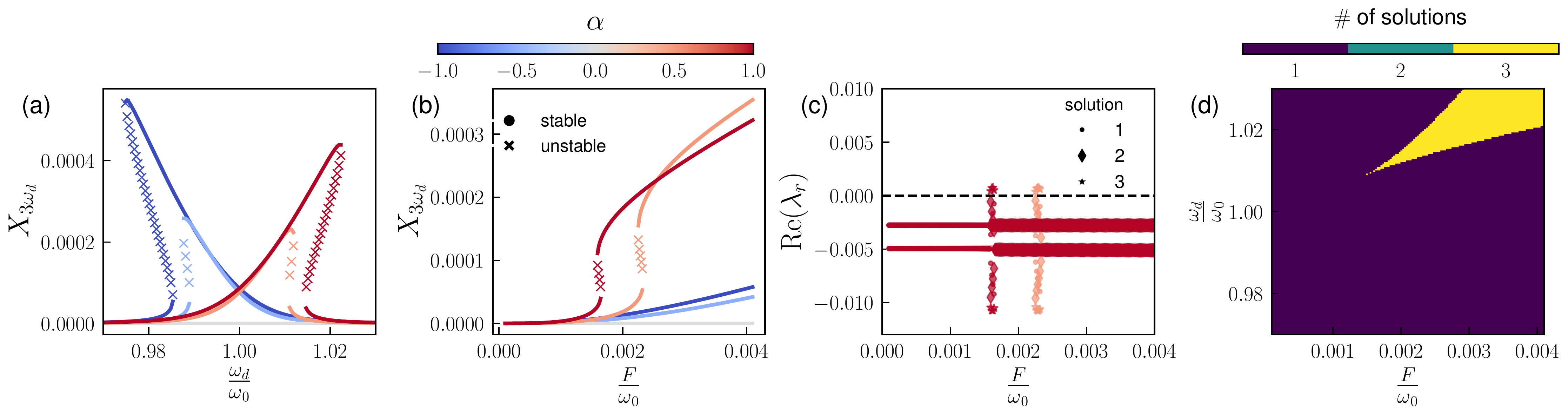}}

       \caption{\textbf{Steady-state solutions for a single-frequency-driven Duffing resonator.}  (top) Harmonic amplitudes as a function of (a) the driving frequency ratio $\omega_d/\omega_0$ ($F/\omega_0=0.0025$) and (b) driving amplitude ratio $F/\omega_0$ (positive detuning, $\omega_d=1.01\omega_0$). Multiple values of $\alpha\in(-1,1)$ are color-encoded. (c) Real parts of complex eigenvalues of Jacobian matrix for solutions in (b). Unstable solutions show positive values for at least one of the eigenvalues. (d) Two-dimensional phase diagram depicting the number of steady-state solutions in the $\{F/\omega_0,\omega_d/\omega_0\}$ plane for $\alpha=1$. Panels at the top (bottom) row correspond to calculation including a single (two) relevant harmonic(s) and plot $X_{\omega_d}$ ($X_{3\omega_d}$). For all panels, $\gamma/\omega_0=0.01$.}
       \label{fig:duffing}
    \end{figure} 
	
	The simplest use case for HarmonicBalance.jl is a driven Duffing resonator governed by Eq.~\eqref{eq:duff_basic}.
	For a driving frequency $\omega_d$ in the vicinity of the natural resonance frequency $\omega_0$, the dominant behaviour can be captured by a single harmonic, the excitation frequency $\omega_d$. The basic lines of code follow below: definition of the system, implementation of the harmonic ansatz 
	\begin{equation}
	    x(t) = u_1(T) \cos(\omega_dt ) + v_2(T) \sin(\omega_dt)\,,
	\end{equation}
	and derivation of corresponding harmonic equations proceed as
    \begin{lstlisting}[numbers=none]
    using HarmonicBalance
    @variables α, ωd, ω0, F, t, γ, x(t) # declare constant variables and a function x(t)
    diff_eq = DifferentialEquation(d(d(x, t),t) + ω0^2 * x + γ*d(x,t)  + α*x^3 ~ F*cos(ωd*t), x)
    add_harmonic!(diff_eq, x, ωd) # specify the ansatz x = u(T) cos(ωdt) + v(T) sin(ωdt) 
    # implement ansatz to get harmonic equations
    harmonic_eq = get_harmonic_equations(diff_eq)
    \end{lstlisting}
    \texttt{harmonic\_eq} is a \texttt{HarmonicEquation} object, corresponding to Eq.~\eqref{eq:duff1_harmeq} in Appendix~\ref{sec:slow_dif}. The output of this code is
    \begin{lstlisting}[numbers=none, basicstyle=\scriptsize\ttfamily]
        A set of 2 harmonic equations
        Variables: u1(T), v1(T)
        Parameters: α, ω, γ, ω0, F
        
        Harmonic ansatz: 
        x(t) = u1*cos(ωd*t) + v1*sin(ωd*t)
        
        Harmonic equations:
        
        (ω0^2)*u1(T) + γ*Differential(T)(u1(T)) + (3//4)*α*(u1(T)^3) + (2//1)*ω*Differential(T)(v1(T)) + γ*ω*v1(T) + (3//4)*α*(v1(T)^2)*u1(T) - F - (ω^2)*u1(T) ~ 0
        
        γ*Differential(T)(v1(T)) + (ω0^2)*v1(T) + (3//4)*α*(v1(T)^3) + (3//4)*α*(u1(T)^2)*v1(T) - (2//1)*ω*Differential(T)(u1(T)) - (ω^2)*v1(T) - γ*ω*u1(T) ~ 0
    \end{lstlisting}
    To calculate the solutions for multiple values of $\omega_d$, we introduce the tuples
    \begin{lstlisting}[numbers=none]
    fixed = (α => 1., ω0 => 1.0, F => 0.0025, γ=> 0.01) # fixed parameters
    swept = ωd => LinRange(0.97, 1.03, 100) # range of parameter values
    \end{lstlisting}
    and simply call
    \begin{lstlisting}[numbers=none]
    solutions = get_steady_states(harmonic_eq, swept, fixed)
    \end{lstlisting}
    The steady-state amplitude corresponding to the harmonic $\omega_d$ is given by $X_{\omega_d}=\sqrt{u_1^2+v_1}$; this can be inspected via
    \begin{lstlisting}[numbers=none]
    plot_1D_solutions(solutions, x="ωd", y="sqrt(u1^2 + v1^2)")
    \end{lstlisting}
    which produces curves similar to Figs.~\ref{fig:duffing}a,b (top row). In such curves, markers are employed to denote the stability of solutions, determined from the Jacobian matrix eigenvalues, which are evaluated at each point [Fig.~\ref{fig:duffing}c]. Real and imaginary parts of Jacobian eigenvalues are displayed via a call to 
   \begin{lstlisting}[numbers=none]
    plot_1D_jacobian_eigenvalues(soln_1d, x="ωd");
    \end{lstlisting}
    Moreover, simultaneous parametric sweeps over many parameter dimensions can be produced by adding ranges to \texttt{swept}, e.g., to get a 2D dataset,
    \begin{lstlisting}[numbers=none]
    swept = (F =>LinRange(0.0001,0.0041,150), ωd=>LinRange(0.97, 1.03,150))
    \end{lstlisting}
    A 2D phase diagram depicting the total number of solutions, irrespective of stability [Fig.~\ref{fig:duffing}d], can then be produced by
    \begin{lstlisting}[numbers=none]
    plot_2D_phase_diagram(solutions,stable=false,observable="nsols")
    \end{lstlisting}
    
    The cubic nonlinearity leads to multiple frequency upconversion processes (Section~\ref{sec:harm_exp}). In particular, we focus on the conversion of a resonantly excited oscillation at $\omega_d\approx\omega_0$ to an off-resonant oscillation at $3\omega_d$\footnote{The Duffing nonlinearity can also enact the conversion of a response excited below resonance at $\omega_d\approx\omega_0/3$, to a resonant oscillation at $3\omega_d\approx\omega_0$.}. The analysis of this mechanism is helpful to study convergence studies of the ansatz. To corroborate it, we implement the ansatz 
    \begin{equation}
        x(t) = u_1(T) \cos(\omega_dt ) + v_1(T) \sin(\omega_dt)+ u_2(T) \cos(3\omega_dt ) + v_2(T) \sin(3\omega_dt)\,,
    \end{equation}
    including harmonics at $\omega_d$ and $3\omega_d$:
    \begin{lstlisting}[numbers=none]
    add_harmonic!(diff_eq, x, ωd) 
    add_harmonic!(diff_eq, x, 3*ωd) 
    harmonic_eq = get_harmonic_equations(diff_eq)
    \end{lstlisting}
    The resulting \texttt{HarmonicEquation} object is printed out
    \begin{lstlisting}[numbers=none, basicstyle=\scriptsize\ttfamily]
        A set of 4 harmonic equations
        Variables: u1(T), v1(T), u2(T), v2(T)
        Parameters: ω, ω0, α, γ, F
        
        Harmonic ansatz: 
        x(t) = u1*cos(ωt) + v1*sin(ωt) + u2*cos(3ωt) + v2*sin(3ωt)
        
        Harmonic equations:
        
        γ*Differential(T)(u1(T)) + (ω0^2)*u1(T) + (3//4)*α*(u1(T)^3) + γ*ω*v1(T) + (2//1)*ω*Differential(T)(v1(T)) + (3//4)*α*(v1(T)^2)*u1(T) + (3//2)*α*(u2(T)^2)*u1(T) + (3//2)*α*(v2(T)^2)*u1(T) + (3//4)*α*(u1(T)^2)*u2(T) + (3//2)*α*u1(T)*v1(T)*v2(T) - F - (ω^2)*u1(T) - (3//4)*α*(v1(T)^2)*u2(T) ~ 0
        
        γ*Differential(T)(v1(T)) + (ω0^2)*v1(T) + (3//4)*α*(v1(T)^3) + (3//4)*α*(u1(T)^2)*v1(T) + (3//4)*α*(u1(T)^2)*v2(T) + (3//2)*α*(u2(T)^2)*v1(T) + (3//2)*α*(v2(T)^2)*v1(T) - (2//1)*ω*Differential(T)(u1(T)) - (ω^2)*v1(T) - γ*ω*u1(T) - (3//4)*α*(v1(T)^2)*v2(T) - (3//2)*α*u1(T)*u2(T)*v1(T) ~ 0
        
        γ*Differential(T)(u2(T)) + (ω0^2)*u2(T) + (1//4)*α*(u1(T)^3) + (3//4)*α*(u2(T)^3) + (6//1)*ω*Differential(T)(v2(T)) + (3//4)*α*(v2(T)^2)*u2(T) + (3//1)*γ*ω*v2(T) + (3//2)*α*(u1(T)^2)*u2(T) + (3//2)*α*(v1(T)^2)*u2(T) - (9//1)*(ω^2)*u2(T) - (3//4)*α*(v1(T)^2)*u1(T) ~ 0
        
        (ω0^2)*v2(T) + γ*Differential(T)(v2(T)) + (3//4)*α*(v2(T)^3) + (3//4)*α*(u1(T)^2)*v1(T) + (3//2)*α*(u1(T)^2)*v2(T) + (3//4)*α*(u2(T)^2)*v2(T) + (3//2)*α*(v1(T)^2)*v2(T) - (1//4)*α*(v1(T)^3) - (6//1)*ω*Differential(T)(u2(T)) - (9//1)*(ω^2)*v2(T) - (3//1)*γ*ω*u2(T) ~ 0
    \end{lstlisting}
    Steady states can now be retrieved again from
    \begin{lstlisting}[numbers=none]
    solutions = get_steady_states(harmonic_eq, swept, fixed)
    \end{lstlisting}
      In Fig.~\ref{fig:duffing} bottom, the nonlinearity, yields a finite amplitude $X_{3\omega_d}=\sqrt{u_2^2+v_2^2}>0$ for the third harmonic $3\omega_d=\omega_0$. Comparison of phase diagrams for the two ansatzes in ~Fig.~\ref{fig:duffing}d does not reveal new regions in terms of the number of solutions. As described in Appendix~\ref{sec:slow_dif}, this insight justifies a perturbative treatment of the upconverted response.
	
	\subsection{Navigating involved solution landscapes: Coupled Duffing resonators}\label{sec:coup_duffing}
	  The power of the homotopy continuation method lies in its ability to capture all possible solutions as parameters are varied. In contrast, experiments in nonlinear systems often implement parameter variations by adiabatic parameter sweeps, which can only access a subset of steady states. This is because such protocol reaches only one steady state at a time. 
	  
	  The knowledge of all possible solutions can be beneficial to interpret experiments of coupled nonlinear oscillators. As the number of oscillators increases, so does the dimensionality of the problem and the number of potential steady states (Section~\ref{sec:harmonic_eq_solving}) with corresponding intricate solution topology. As an example of how increasingly complex problems are suitable for treatment by HarmonicBalance.jl, we consider in this section the harmonic equations for two coupled Duffing resonators (natural frequencies $\omega_{x}$ and $\omega_y$, coupling $k$), governed by
	  \begin{align}
	      &\ddot{x}(t)+\omega_{x}^{2}x(t)+\gamma\dot{x}(t)+\alpha x(t)^{3}-ky(t)=F \cos(\omega_d t),\\\
	      &\ddot{y}(t)+\omega_{y}^{2}y(t)+\gamma\dot{y}(t)+\alpha y(t)^{3}-kx(t)=F \cos(\omega_d t).
	  \end{align}
	  In addition to obtaining steady states for multiple parameters using homotopy continuation, we solve the time dynamics to simulate an experimentally standard parametric sweep via an interface with \texttt{DifferentialEquations.jl}. This can be attained via the following lines of code
	  \begin{lstlisting}[numbers=none]
      @variables ω_x, ω_y, t, ω, F, γ, α, k, x(t), y(t);
      
      free_eq = [ d(d(x, t),t) + ω_x^2 * x +γ*d(x,t) + α*x^3 - k*y, 
        d(d(y,t),t) + ω_y^2*y + γ*d(y,t) + α*y^3 - k*x]

      forces = [F*cos(ω*t), F*cos(ω*t)]

      diff_eq = DifferentialEquation(free_eq - forces, [x, y])

      add_harmonic!(diff_eq, x, ω) # x will oscillate at ω
      add_harmonic!(diff_eq, y, ω) # y will oscillate at ω
      harmonic_eq = get_harmonic_equations(diff_eq)
    \end{lstlisting}
    Execution of the above block also prints a detailed summary of the simplified system
    \begin{lstlisting}[numbers=none, basicstyle=\scriptsize\ttfamily]
        A set of 4 harmonic equations
        Variables: u1(T), v1(T), u2(T), v2(T)
        Parameters: α, ω, ω_x, γ, k, F, ω_y
        
        Harmonic ansatz: 
        x(t) = u1*cos(ωt) + v1*sin(ωt)
        y(t) = u2*cos(ωt) + v2*sin(ωt)
        
        Harmonic equations:
        
        (ω_x^2)*u1(T) + γ*Differential(T)(u1(T)) + (3//4)*α*(u1(T)^3) + (2//1)*ω*Differential(T)(v1(T)) + γ*ω*v1(T) + (3//4)*α*(v1(T)^2)*u1(T) - F - k*u2(T) - (ω^2)*u1(T) ~ 0
        
        γ*Differential(T)(v1(T)) + (ω_x^2)*v1(T) + (3//4)*α*(v1(T)^3) + (3//4)*α*(u1(T)^2)*v1(T) - (ω^2)*v1(T) - k*v2(T) - (2//1)*ω*Differential(T)(u1(T)) - γ*ω*u1(T) ~ 0
        
        γ*Differential(T)(u2(T)) + (ω_y^2)*u2(T) + (3//4)*α*(u2(T)^3) + (2//1)*ω*Differential(T)(v2(T)) + γ*ω*v2(T) + (3//4)*α*(v2(T)^2)*u2(T) - F - k*u1(T) - (ω^2)*u2(T) ~ 0
        
        (ω_y^2)*v2(T) + γ*Differential(T)(v2(T)) + (3//4)*α*(v2(T)^3) + (3//4)*α*(u2(T)^2)*v2(T) - k*v1(T) - (ω^2)*v2(T) - (2//1)*ω*Differential(T)(u2(T)) - γ*ω*u2(T) ~ 0
    \end{lstlisting}
    Multi-parameter steady states can be obtained,
    \begin{lstlisting}[numbers=none]
     fixed_parameters = (ω_x=> 1., ω_y=> 1.05, γ=> 2E-3, F=> 1E-2, α=> 1E-1, k=> 5E-2])
     sweep = (ω => LinRange(0.85,1.25,150))
     solutions = get_steady_states(problem, sweep, fixed)
	 \end{lstlisting}
	 This yields a \texttt{Result} object,
	 \begin{lstlisting}[numbers=none, basicstyle=\scriptsize\ttfamily]
    	   Solution branches:   11
           of which real:    7
           of which stable:  4
      \end{lstlisting}
	  We visualise the amplitudes of stable steady-state solutions via
	  \begin{lstlisting}[numbers=none]
      plot_1D_solutions(soln, x="ω", y="sqrt(u1^2 + v1^2)", plot_only=["physical", "stable"],filename="coupled_duffing_sols");
	  \end{lstlisting}
	  Here we saved results to a $\texttt{.jld2}$ file by filling the \texttt{filename} keyword argument. Data can be subsequently recovered via use of the \texttt{load} function. 
	  
\begin{figure}[t]
        \centering
        \includegraphics[width=0.7\textwidth]{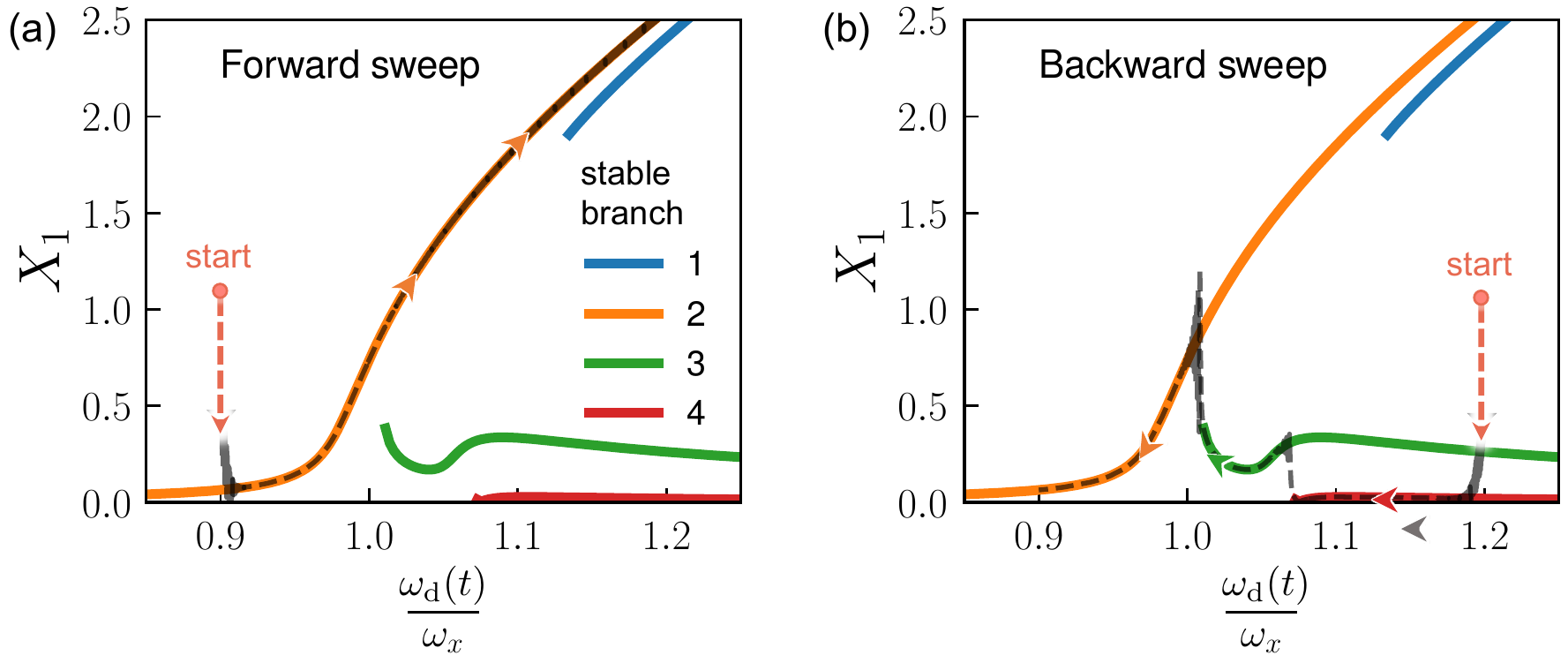}
        \caption{\textbf{Time-dependent driving frequency sweep (gray) and stable steady-state landscape (colored) in two linearly coupled, driven Duffing resonators.} Steady-state solution amplitudes $X_1=\sqrt{u_1^2+v_1^2}$
        for a system initiated away from any steady state, rapidly collapsing into one of the available solutions. Adiabatic variation of the driving frequency follows a subset of stable, steady states. The arrows indicate the direction of the sweep; jumps occur at points where a solution branch ceases to be stable. The observed behaviour is hysteretic, i.e., it depends on the sweep direction. The parameters chosen are $\omega_{x}=1,\omega_{y}=1.05$, $\gamma=2\cdot 10^{-3}$, $F=10^{-2}$, $\alpha=10^{-1}$, $J=5\cdot10^{-2}$. The sweeping time is $\tau=10^5\gg\gamma^{-1}$, much slower than any relaxation time to minimise non-adiabatic effects.}
        \label{fig:coup_duffing_hyst}
    \end{figure}

	  We are now in a position to check the accessibility of the steady states by performing a backward, time-dependent sweep of the frequency, initiating at a given particular condition:
	  \begin{lstlisting}[numbers=none]
	      # select a solution and evolve from it
            s1 = get_single_solution(soln, branch=5, index=75);

            sweep = HarmonicBalance.TimeEvolution.ParameterSweep(ω => (1.2, 0.9), (0, 1E5)) #HarmonicBalance.TimeEvolution.ParameterSweep(ω => (s1[ω], 0.95), (0, 1E5))
            
            function t_solve(s) """retrieve long-time evolved results from the slow-flow equation"""
                problem = HarmonicBalance.TimeEvolution.ODEProblem(averagedEOM, steady_solution=s, timespan=(0,1E5), sweep=sweep)
                time_soln = HarmonicBalance.TimeEvolution.solve(problem,saveat=10);
            end
            time_soln_1 = t_solve(s1);
	  \end{lstlisting}  
    A forward sweep can also be performed by setting
    \begin{lstlisting}[numbers=none]
        sweep = HarmonicBalance.TimeEvolution.ParameterSweep(ω => (0.9,1.2), (0, 1E5))
        time_soln_2 = t_solve(s1);
    \end{lstlisting}
    Large-time amplitudes from backwards/forward time sweeps can be retrieved from
    \begin{lstlisting}[numbers=none]
    X1 = sqrt.(getindex.(time_soln_1.u,1).^2 .+ getindex.(time_soln_1.u,2).^2)
    X2 = sqrt.(getindex.(time_soln_2.u,1).^2 .+ getindex.(time_soln_2.u,2).^2)
    \end{lstlisting}
    %
    %or from the \texttt{transform\_solutions} method.
     Results from these sweeps are presented along with the steady states in Fig.~\ref{fig:coup_duffing_hyst}.
       
	\begin{figure}[t]
        \centering
        \includegraphics[width=1\textwidth]{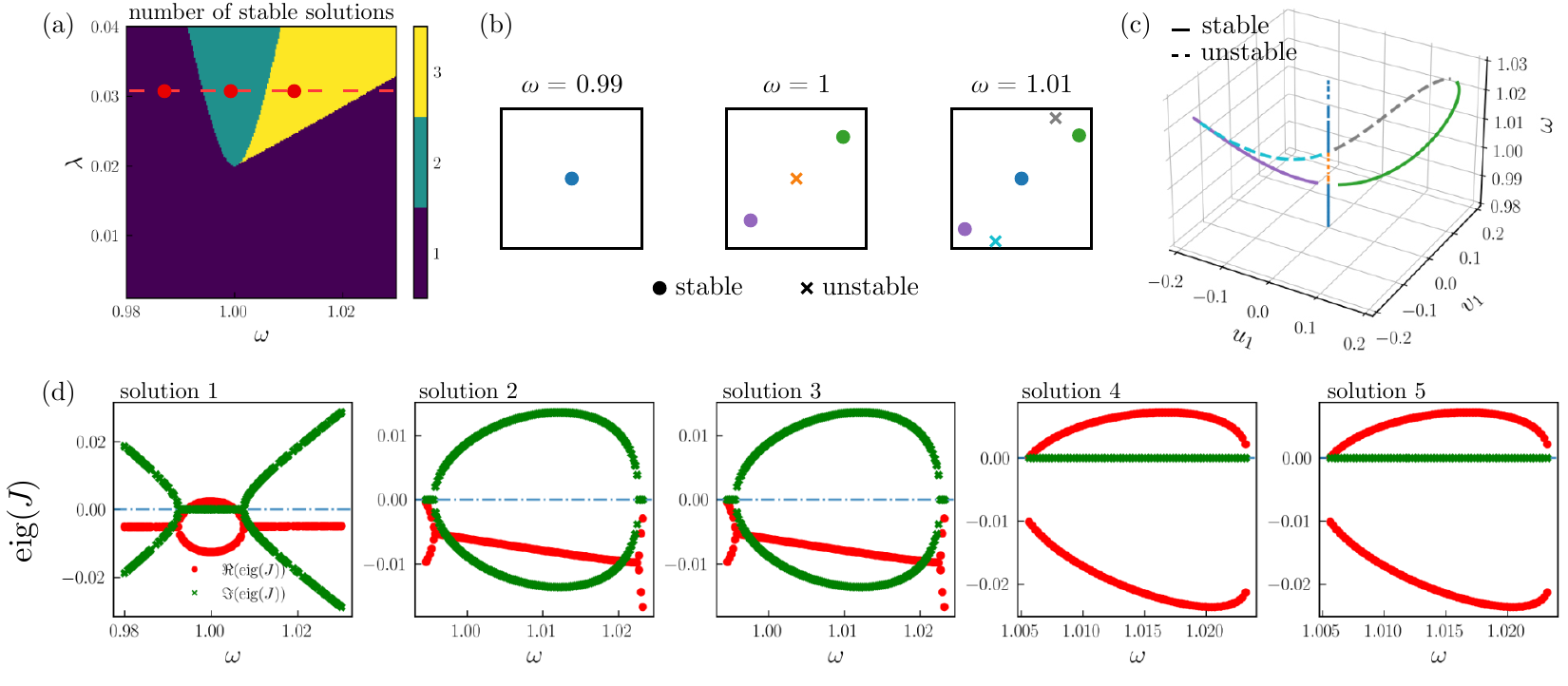}
        \caption{(a) Two-dimensional phase diagram depicting the number of steady-state solutions in the $\{\lambda,\omega_d\}$ plane. (b) Representative steady-state solutions depicted in the $u-v$ plane as indicated in (a). (c) Steady-state solutions as a function of driving frequency $\omega_d$. (d) Eigenvalues of the Jacobian [Eq.~\eqref{eq:Jacobeq}] evaluated for the solutions in (c). Other parameter values are $\alpha=1$, $\gamma = 0.02$, and $\eta = 0.3$.}
        \label{fig:parametric}
    \end{figure}

	\subsection{Parametrically driven Duffing oscillator}
	Nonlinear effects in oscillating systems manifest not only when systems are driven through external forcing, as in Secs.~\ref{sec:duffing_example}\ref{sec:coup_duffing} but also through periodic modulation of the system parameters, also known as parametric driving~\cite{Landau_Lifshitz, Lifshitz_2008}. The simplest example of such an oscillator is the parametrically-driven Duffing oscillator or \textit{parametron}~\cite{von1957non,Goto1959,Sterzer1959,Hosoya1991,Mahboob2008}, governed by the damped nonlinear Mathieu equation
	\begin{equation}
	    \ddot{x} +\gamma \dot{x}+ \omega_0^2(1-\lambda\cos(\omega_{\rm p} t))x + \alpha x^3 +\eta x^2 \dot{x}= 0 \:.
	\end{equation}
	%
	%describe double period bifurcation
	Interestingly, this system hosts finite oscillations when the parametric driving frequency approaches  $\omega_{\rm p} \approx \omega_0$. In addition, above a critical driving rate, such a system is known to host two stable oscillation states with equal amplitudes but $\pi$-shifted phases. Their (subharmonic) response frequency, however, is at $\omega = \omega_{\rm p}/2$. We thus study the response at frequency $\omega$ via the following lines of code:
	\begin{lstlisting}[numbers=none]
	@variables ω0,γ,λ,ωp,ψ,α,η,t,T,x(t)
    free_eq =  d(d(x,t),t) + γ*d(x,t) + ω0^2*(1-λ*cos(2*ω*t+ψ))*x + α * x^3 + η *d(x,t) * x^2
    diff_eq = DifferentialEquation(free_eq, x)
    add_harmonic!(diff_eq, x, ω);
    harmonic_eq = get_harmonic_equations(diff_eq)
	\end{lstlisting}
	We can now study the 2D-phase diagram spanned by $\lambda$ and $\omega$ (Fig.~\ref{fig:parametric}a). For it, we define the fixed parameters and specify the parameter range for $\lambda$ and $\omega$:
	\begin{lstlisting}[numbers=none]
	fixed_parameters = ParameterList(ω0 => 1.0,γ => 1E-2, λ => 3E-2, F => 0,  α => 1.,  η=>0.3, θ => 0, ψ => 0)
    swept_parameters = ParameterRange(ω => collect(LinRange(0.98, 1.03, 200)), λ => collect(LinRange(0.001,0.04,200)))
    phase_diagram = get_steady_states(problem, swept_parameters, fixed_parameters; random_warmup=false, threading=false, sorting="nearest")
	\end{lstlisting}
	We can now plot the phase diagram via
	\begin{lstlisting}[numbers=none]
	plot_2D_phase_diagram(phase_diagram, stable=false,observable="binary")
	\end{lstlisting}
	Depending on the argument of \texttt{observable} the phase diagram is either characterised by the number of solutions or the type and stability of solutions found for specific parameter configurations (Section~\ref{subsec:vis}). In this problem, they both yield the same result. Representative steady solutions of the different phases can be studied in the phase plane spanned by $(u,v)$. The bifurcation points at the phase boundaries are  studied best by using 3D-plot showing the $(u,v)$ representation as a function of a system parameter, e.g. $\omega$ [see Fig.~\ref{fig:parametric}(c)]. They can be obtained using the function $\texttt{plot\_1D\_solutions\_spaghetti}$.
	
	Important insights for the stability of the solutions and their bifurcation can be gained by studying the eigenvalues of the Jacobian matrix [see Eq.~\eqref{eq:Jacobeq}] (Fig.~\ref{fig:parametric}d. They can be easily calculated from the steady state solutions using the function $\texttt{plot\_1D\_jacobian\_eigenvalues}$. Note that such a representation reveals the transition from the region with a single stable solution to the region with two solutions corresponding to a pitchfork bifurcation, in the vicinity of which the imaginary part vanishes, while the real part splits~\cite{Soriente2021}. %Additional interesting behaviour can also be found when the system is close to the instability lobes...

	\subsection{Increasing computational complexity: Performance scaling}\label{sec:hardexample}
    Here, we shortly illustrate the performance of HarmonicBalance.jl for varying system sizes. We consider a chain of linearly coupled Duffing oscillators, each with nonlinear damping (amplitude~$\eta$, other parameters defined above Eq.~\eqref{eq:duffingFT}),
    \begin{equation} \label{eq:Duffing_chain}
    \ddot{x}_i(t) + \omega_0^2 x_i(t) +  \alpha x_i(t)^3 + \eta x_i(t)^2 \dot{x}_i(t) - k \sum_{j=i\pm1} x_j(t) = F \cos(\omega t)\:, \quad i = 1,2,..., N\:.
    \end{equation}
     Similar systems have been explored in the context of combinatorial optimisation machines based on the mapping of effective spins to parametron networks~\cite{Wang2013,Bello2019a,CalvaneseStrinati2019,2019PhRvL.123y4102F,CalvaneseStrinati2020,PhysRevResearch.4.013149}.
    The displacement of each oscillator, $x_i$, is expanded using the harmonic oscillating at $\omega$, giving the harmonic ansatz
    \begin{equation}
    x_i(t) = u_i(T) \cos(\omega t ) + v_i(T) \sin(\omega t).
    \end{equation}
    As explained in Section~\ref{sec:model}, each oscillator adds 2 harmonic equations of order 3. Therefore, a chain of length $N$ and $M=1$ leads to a set of $2{N}$ equations. The corresponding B\'{e}zout bound on the number of solutions, and hence the number of paths which must be tracked by the homotopy continuation algorithm, is $3^{2{N}}$. 
    
    In Table.~\ref{table:benchmark}, we show the computational times necessary for symbolic manipulation and homotopy solving as well as the number of complex and real solutions found. We highlight solving a chain of $N=5$ resonators still takes a few minutes on a single CPU. Notably, the number of unique solutions is significantly smaller than the B\'{e}zout bound in all cases, with the number of real solutions being smaller still. The bottleneck is, therefore, the initial tracking of the homotopy from $3^{2{N}}$ paths to the relatively few non-singular ones. Subsequent steps such as solving for multiple parameter sets (i.e. tracking parameter homotopy paths) and determining stability (Jacobian evaluation and diagonalisation) only involve non-singular paths and thus are relatively inexpensive.
    
	\begin{table}[t] 
	    \centering
        \begin{tabular}{ |c|c|c|c|c|c|} 
         \hline
         $N$ & 1 & 2 & 3 & 4 & 5 \\ \hline
         $t_{\mathrm{symbolic}}$ [s] & 0.16 & 0.38 & 0.71 & 1.11 & 1.54 \\
         $t_{\mathrm{solve}}$ [s] & 0.52 & 1.47 & 17.3 & 173 & 1801 \\
         B\'{e}zout bound & 9 & 81 & 729 & 6561 & 59049 \\
         complex solutions & 3 & 11 & 59 & 545 & 3577\\
         real solutions & 3 & 11 & 32 & 103 & 310 \\
         \hline
        \end{tabular}
	    \caption{\textbf{Finding the steady states of Eq.~\eqref{eq:Duffing_chain} for varying chain length $N$.} The rows describe (i)~Computational time required to obtain the harmonic equations and the symbolic Jacobian $t_{\mathrm{symbolic}}$. (ii)~Time to solve the harmonic equations for 50 parameter sets $t_{\mathrm{solve}}$. (iii)~The B\'{e}zout bound. (iv)~Number of complex solutions found. (v)~Number of real solutions found. The parameters used are $\omega_0 = 1, \alpha = 1, k=0.1, \eta = 0.1, F=0.01$. A single core of an Intel i7-8550U CPU was used.}
	    \label{table:benchmark}
	\end{table}

    Two remarks regarding performance are in order. First, the process of path tracking is naturally well-suited for parallelisation, and HomotopyContinuation.jl does include the necessary functionality. Second, systems such as the Duffing chain possess spatial symmetries (in this case, inversion symmetry) and internal/gauge symmetries (e.g. discrete time translation, $u_i\mapsto -v_i, v_i\mapsto u_i$) which cause some solutions to be degenerate. Making use of this property can further reduce the computational time needed.

	\section{Conclusion}
	In this work, we have introduced a Julia package to treat the steady-state behaviour of generic nonlinear, polynomial, dynamical systems with harmonic time dependence. Combining symbolic and numerical calculations with simple graphical capabilities, our package is developed with simplicity of use in mind while enabling the study of complex nonlinear systems.
	
	Its modular design paves the way for future methodological extensions, including detection of Hopf bifurcations~\cite{veltz:hal-02902346}, limit cycles and chaos~\cite{Genesio1992}, higher-order Krylov-Bogoliubov averaging method~\cite{krylov2016introduction}, as well as interfaces with existing dedicated libraries to treat nonlinear spatially-extended or quantum systems~\cite{kramer2018quantumoptics,QuantumAlgebra.jl}. Its usage can assist a breadth of fields, where nonlinear harmonically-driven systems appear, such as modal analysis in structural dynamics~\cite{Ewins2000,Kerschen2006}, electric circuits~\cite{Rohde2005,Rubiola2008,Fallis2003}, nonlinear optics~\cite{Haken1975,Shen2002,del2007optical,Rodriguez2016,DelPino2016,Sounas2018,Zambon_2020,peters2021limit}, optomechanics~\cite{Aspelmeyer2014,Pelka2020,Burgwal2020,Roque2020}, micro- and nanomechanics~\cite{Poot2012,Papariello2016,Yang_2021a, Yang_2021b, Mohammadi2020,Huber_2020,2018ApPhL.112w3105E,2020PhRvP..14a4042K,2021arXiv210911943H}, oscillator networks~\cite{2016PhRvA..93d3827O,2019PhRvL.123l4301H,Ozawa2019,2021Kosata,del2021non,2022arXiv220106315P},  Ising machines~\cite{Wang2013,Bello2019a,CalvaneseStrinati2019,2019PhRvL.123y4102F,CalvaneseStrinati2020,PhysRevResearch.4.013149}, and many-body light-matter systems~\cite{griffin1996bose,Carusotto2013,Soriente2018,Kirton2019,Soriente2020,Soriente2021,2021PhRvX..11d1046F}. 
    
    The ability to explore the complete solution landscape with an ansatz that has a tunable level of complexity makes our tool ideal for working alongside experiments. Specifically, high-order processes (e.g., up-conversion) can be readily accounted for through the progressive addition of harmonic frequencies. We hope that the free availability of a user-friendly, high-performance code for harmonic nonlinear dynamics calculations will help advance multiple disciplines by allowing researchers to perform computations currently considered out of reach due to their complexity. In addition, we hope its implementation, sitting on Julia's rich ecosystem for high-performance scientific computing, will help nurture a multidisciplinary open-source community.    
    
	\section*{Acknowledgements}

	\paragraph{Author contributions}
	J.K. and J.d.P. developed the package, T.H. provided insight on the theoretical approach, and O.Z. supervised the project. All authors contributed to the interpretation of results and writing of the manuscript.
    
    \paragraph{Funding information}
    
    J.K., T.H. were supported by the Swiss National Science Foundation through grant CRSII5 $177198/1$. J.d.P. was supported by the ETH Fellowship program (grant no. 20-2 FEL-66). O.Z.  acknowledges funding through SNSF grant PP00P2\_190078 and from the Deutsche Forschungsgemeinschaft (DFG) - project number 449653034.
    
	\appendix
	
    % format the equation environment
    \renewcommand{\theequation}{A.\arabic{equation}}
    % reset the counter
    \setcounter{equation}{0}
    
	\section{Harmonic equations for a Duffing resonator} \label{sec:slow_dif}
	
	Here, we derive the harmonic equations for a single Duffing resonator, governed by the equation
	\begin{equation} \label{eq:duff_ex}
	    \ddot{x}(t) + \omega_0^2 x(t) + \alpha x^3(t) = F \cos(\omega_d t + \theta)\:, 
	\end{equation}
	where we considered driving with a finite phase offset $\theta$ for generality.
	As explained in Section~\ref{sec:model}, for a periodic driving at frequency $\omega_d$ and a weak nonlinearity $\alpha$, we expect the response at frequency $\omega_d$ to be dominant, followed by a response at $3\omega_d$ due to frequency conversion.
	
	\subsubsection*{Expanding in $\boldsymbol{\omega_d}$ only} 
	We first describe the established~\cite{Nayfeh_2008, Rand_2005, Lifshitz_2008} approach to finding the steady states of Eq.~\eqref{eq:duff_ex}, where frequency conversion is only added perturbatively. The starting point is a harmonic ansatz for $x$ of the form Eq.~\eqref{eq:ansatz}, containing a single frequency $\omega_d$,
	\begin{equation} \label{eq:duff_1harm}
	    x(t) = u(T) \cos(\omega_d t) + v(T) \sin(\omega_d t)\:,
	\end{equation}
	with the harmonic variables $u$ and $v$. The \textit{slow time} $T$ is, for now, equivalent to $t$. Substituting this ansatz into Eq.~\eqref{eq:duff_ex} results in
	\begin{align} \label{eq:duff1_eingesetzt}
	\left[\ddot{u} + 2 \omega_d \dot{v} + u \left(\omega_0^2 - \omega_d^2 \right) +  \frac{3 \alpha \left(u^3 + uv^2\right)}{4} + F \cos{\theta}\right] &\cos(\omega_d t)& \nonumber\\
	+ \left[\ddot{v} - 2 \omega_d \dot{u} + v \left(\omega_0^2 - \omega_d^2 \right)  +\frac{3 \alpha \left(v^3 + u^2 v\right)}{4} - F \sin{\theta}\right] &\sin(\omega_d t)& \nonumber\\
	+ \frac{\alpha \left(u^3 - 3 u v^2\right)}{4} \cos(3 \omega_d t) +  \frac{\alpha \left(3u^2 v - v^3\right)}{4} \sin(3 \omega_d t) &= 0.
	\end{align}
	
	We see that $x^3$ in Eq.~\eqref{eq:duff_ex} generates terms that oscillate at $3\omega_d$, describing the process of frequency upconversion. We now Fourier-transform both sides of Eq.~\eqref{eq:duff1_eingesetzt} with respect to $\omega_d$ to obtain the harmonic equations. This process is equivalent to extracting the respective coefficients of $\cos(\omega_d t)$ and $\sin(\omega_d t)$. Here the distinction between $t$ and $T$ becomes important: since the evolution of $u(T)$ and $v(T)$ is assumed to be slow, they are treated as constant for the purpose of Fourier transformation. Since we are interested in steady states, we drop the higher-order derivatives and rearrange the resulting equation to the form of Eq.~\eqref{eq:harmoniceq} of the main text 
	\begin{equation} \label{eq:duff1_harmeq}
	\frac{d}{dT} \mqty(u \\ v ) = \frac{1}{8 \omega_d} \mqty(4 v \left(\omega_0^2-\omega_d^2 \right) + 3 \alpha \left(v^3 + u^2 v  \right) - 4 F \sin{\theta}  \\ 4 u \left(\omega_d^2-\omega_0^2 \right)  - 3 \alpha \left(u^3 + u v^2 \right) - 4 F \cos{\theta}  ) \;.
	\end{equation}
	Note that our assumption that $u(T)$ and $v(T)$ are slowly changing, i.e. composed of small frequency terms also sets constraints in Fourier space: we neglect all the frequencies which are not close to $\omega_d$. In the extreme case of constant $u$ and $v$, the described frequency range reduces to the discrete frequency $\omega_d$.
	Steady states can now be found by setting the l.h.s. to zero, i.e., assuming $u(T)$ and $v(T)$ constant and neglecting any transient behaviour. This step is referred to in the literature as "balancing the harmonics"~\cite{Krack_2019}. This results in a set of 2 nonlinear polynomial equations of order 3, for which the maximum number of solutions set by B\'{e}zout theorem is $3^2=9$. Depending on the parameters, the number of real solutions is known to be between 1 and 3, see the solution diagrams Section~\ref{sec:examples}.
	
	The steady states describe a response that may be recast as $x_0(t) = X_0 \cos(\omega_d t + \phi)$, where $X_0=\sqrt{u^2+v^2}$ and $\phi=-\atan(v/u)$. Frequency conversion from $\omega_d$ to $3 \omega_d$ can be found by setting $x(t) \equiv x_0(t) + \delta x(t)$ with $|\delta x(t)|\ll|x_0(t)|$ and expanding Eq.~\eqref{eq:duff_ex} to first-order in $\delta x(t)$. The resulting equation
	\begin{equation}
	    \delta \ddot{x}(t) + \left[\omega_0^2 + \frac{3 \alpha X_0^2}{4} \right]\delta x(t) = - \frac{\alpha X_0^3}{4} \cos(3 \omega_d t + 3 \phi)\,,
	\end{equation}
	describes a simple harmonic oscillator, which is exactly soluble. Correspondingly, a response of $\delta x(t)$ at frequency $3 \omega_d$ is observed. Since this response is obtained perturbatively for each steady state of Eq.~\eqref{eq:duff1_harmeq}, no previously-unknown solutions are generated in the process.
	
	\subsubsection*{Expanding in $\boldsymbol{\omega_d}$ and $\boldsymbol{3\omega_d}$}

	An approach in the spirit of harmonic balance is to use both harmonics $\omega_d$ and $3 \omega_d$ on the same footing, i.e., to insert the ansatz
	\begin{equation}
	    x(t) = u_1(T) \cos(\omega_d t) + v_1(T) \sin(\omega_d t) + u_2(T) \cos(3 \omega_d t) + v_2(T) \sin(3\omega_d t)\:,
	\end{equation}
	with $u_1, u_2, v_1, v_2$ being the harmonic variables. As before we substitute the ansatz into Eq.~\eqref{eq:duff_ex}, drop second derivatives with respect to $T$ and Fourier-transform both sides. Now, the respective coefficients correspond to $\cos(\omega_d t)$, $\sin(\omega_d t)$, $\cos(3 \omega_d t)$ and $\sin(3\omega_d t)$. Rearranging, we obtain
	\begin{align} \label{eq:duff_harmeq}
	\begin{split}
	\frac{du_1}{dT} &=  \frac{1}{2\omega_d} \left[ \left({\omega_0}^{2} - \omega_d^2 \right) v_1 + \frac{3\alpha}{4} \left( v_1^3 + u_1^2 v_1 + u_1^2 v_2 - v_1^2 v_2 + 2 u_2^2 v_1 + 2 v_2^2 v_1 - 2 u_1 u_2 v_1\right)  + F \sin{\theta} \right],
	\\
	\frac{dv_1}{dT} &= \frac{1}{2\omega_d} \left[ \left({\omega_d}^{2} - \omega_0^2 \right) {u_1} - \frac{3 \alpha}{4} \left( u_1^3 + u_1^2 u_2 + v_1^2 u_1 - v_1^2 u_2+ 2 u_2^2 u_1 + 2 v_2^2 u_1  + 2 u_1 v_1 v_2\right) - F \cos{\theta} \right],
	\\
	\frac{d u_2}{dT} &= \frac{1}{6 \omega_d} \left[ \left(\omega_0^{2} - 9\omega_d^2 \right) {v_2} + \frac{\alpha}{4} \left( - v_1^3 + 3 v_2^3 + 3 u_1^2 v_1 + 6 u_1^2 v_2 + 3 u_2^2 v_2 + 6 v_1^2 v_2\right) \right],
	\\
	\frac{dv_2}{dT} &= \frac{1}{6 \omega_d} \left[ \left(9\omega_d^2 - \omega_0^2\right) {u_2} - \frac{\alpha}{4} \left( u_1^3 + 3 u_2^3 + 6 u_1^2 u_2 - 3 v_1^2 u_1 + 3 v_2^2 u_2 + 6 v_1^2 u_2\right) \right] \:.
	\end{split}
	\end{align}
    In contrast to Eqs.~\eqref{eq:duff1_harmeq},  we now have 4 equations of order 3, allowing up to $3^4=81$ solutions (the number of unique real ones is again generally far smaller, as seen in Section~\ref{sec:examples}). The larger number of solutions is explained by higher harmonics which cannot be captured perturbatively by the single-frequency ansatz. In particular, those where the $3 \omega_d$ component is significant. Such solutions appear, e.g., for $\omega_d \approx \omega_0 / 3$ where the generated $3 \omega_d$ harmonic is close to the natural resonance frequency.
	
	\section{Solving steady states for multiple parameter values}\label{sec:solving_multiple}
	Evaluating Eq.~\eqref{eq:harmoniceq} at
	each parameter set determines a distinct numerical system of polynomial equations, whose roots can be found using homotopy continuation; this generally requires tracking the maximum number of roots given by B\'{e}zout's theorem, which may be very time-consuming. Having solved for one parameter set, the entire tracking procedure is not necessary for others - one only needs to track paths leading to non-singular solutions. In particular, we carry out
	\begin{enumerate}
	    \item A "warm-up", i.e. path
	tracking from the roots of, e.g., $\vb{U}(\vb{u}_0)=\{u_1^{d_1} - 1,v_1^{d_1} - 1,u_2^{d_2} - 1,v_2^{d_2} - 1,\cdots\}$ ($d_r, r=(1,2,\cdots,NM)$ denotes the degree of the polynomial $\bar{F}_r(\vb{u}_0)$) towards the roots of a "generic" system, namely  $\bar{\vb{F}}(\vb{u}_0)$ evaluated with random complex (unphysical) parameters\footnote{For some classes of polynomials, alternative starting systems (see polyhedral homotopy, implemented in  HomotopyContinuation.jl~\cite{Breiding_2018, Huber_2020}) can significantly decrease the number of solution paths to be tracked compared with a total degree homotopy.}.  All singular paths  are filtered out during this procedure.
	\item The number of complex roots of the generic system is an upper bound for the number of roots of $\bar{\vb{F}}(\vb{u}_0)=0$ for arbitrary parameters~\cite{Sommese2005}. In a final step, we apply a \textit{parameter homotopy} (a mapping between the collection of systems $\bar{\vb{F}}(\vb{u}_0)=0$ obtained by only varying the parameters)  towards a physical real parameter value. This may involve tracking a massively reduced number of paths, hence being much faster than step 1.
	\item Repeat step 2 for each parameter value.
	\end{enumerate}
	
	Despite the computational overhead of the problem's "complexification", complex solution paths allow for a stable path tracking algorithm, preventing singularities that will otherwise show up in a real-space parameter homotopy (see Ref.~\cite{Sommese2005} for details).

	\bibliography{references}
	\nolinenumbers
	
\end{document}